\documentstyle[12pt]{article}
\input{psfig.sty}
\newlength{\dinwidth}
\newlength{\dinmargin}
\newlength{\extraspace}
\newlength{\extraspaces}
\newcommand{\be}{\begin{equation}
\addtolength{\abovedisplayskip}{\extraspaces}
\addtolength{\belowdisplayskip}{\extraspaces}
\addtolength{\abovedisplayshortskip}{\extraspace}
\addtolength{\belowdisplayshortskip}{\extraspace}}
\newcommand{\ee}{\end{equation}}
\newcommand{\bdm}{\begin{displaymath}
\addtolength{\abovedisplayskip}{\extraspaces}
\addtolength{\belowdisplayskip}{\extraspaces}
\addtolength{\abovedisplayshortskip}{\extraspace}
\addtolength{\belowdisplayshortskip}{\extraspace}}
\newcommand{\edm}{\end{displaymath}}
\renewcommand{\thefootnote}{\fnsymbol{footnote}}
\def\simlt{\mathrel{\lower2.5pt\vbox{\lineskip=0pt\baselineskip=0pt
           \hbox{$<$}\hbox{$\sim$}}}}
\def\simgt{\mathrel{\lower2.5pt\vbox{\lineskip=0pt\baselineskip=0pt
           \hbox{$>$}\hbox{$\sim$}}}}
\newcommand{\ls}[1]
   {\dimen0=\fontdimen6\the\font
    \lineskip=#1\dimen0
    \advance\lineskip.5\fontdimen5\the\font
    \advance\lineskip-\dimen0
    \lineskiplimit=.9\lineskip
    \baselineskip=\lineskip
    \advance\baselineskip\dimen0
    \normallineskip\lineskip
    \normallineskiplimit\lineskiplimit
    \normalbaselineskip\baselineskip
    \ignorespaces}

\catcode`@=11
\newcount\@tempcntc
\def\@citex[#1]#2{\if@filesw\immediate\write\@auxout{\string\citation{#2}}\fi
  \@tempcnta\z@\@tempcntb\m@ne\def\@citea{}\@cite{\@for\@citeb:=#2\do
    {\@ifundefined
       {b@\@citeb}{\@citeo\@tempcntb\m@ne\@citea\def\@citea{,}{\bf ?}\@warning
       {Citation `\@citeb' on page \thepage \space undefined}}%
    {\setbox\z@\hbox{\global\@tempcntc0\csname b@\@citeb\endcsname\relax}%
     \ifnum\@tempcntc=\z@ \@citeo\@tempcntb\m@ne
       \@citea\def\@citea{,}\hbox{\csname b@\@citeb\endcsname}%
     \else
      \advance\@tempcntb\@ne
      \ifnum\@tempcntb=\@tempcntc
      \else\advance\@tempcntb\m@ne\@citeo
      \@tempcnta\@tempcntc\@tempcntb\@tempcntc\fi\fi}}\@citeo}{#1}}
\def\@citeo{\ifnum\@tempcnta>\@tempcntb\else\@citea\def\@citea{,}%
  \ifnum\@tempcnta=\@tempcntb\the\@tempcnta\else
   {\advance\@tempcnta\@ne\ifnum\@tempcnta=\@tempcntb \else \def\@citea{--}\fi
    \advance\@tempcnta\m@ne\the\@tempcnta\@citea\the\@tempcntb}\fi\fi}
\catcode`@=12

\newcommand{\prd}{{\em Phys.\ Rev.\ }  {\bf D}}

\newcommand{\prl}{{\em Phys.\ Rev.\ Lett.\ }}
\newcommand{\np}{{\em Nucl.\ Phys.\ }{\bf B}}
\newcommand{\plb}{{\em Phys.\ Lett.\ }{\bf B}}

\newcommand{\zp}{Z. Phys.\ {\bf C}}

\newcommand{\rmp}{{\em Rev.\ Mod.\ Phys.\ }}

\begin{document}
\setcounter{footnote}{1}
\begin{flushright}
MADPH-97-1016\\
October 1997\\
\end{flushright}
\vspace{14mm}
\begin{center}
\Large{{\bf Short Distance 
Coefficients and the Vanishing of the Lepton Asymmetry 
in $B\to V\ell^+\ell^-$}}
\end{center}
\vspace{5mm}
\begin{center}
{
Gustavo Burdman}\footnote{e-mail address: burdman@pheno.physics.wisc.edu}\\
*[3.5mm]
{\normalsize\it Department of Physics, University of Wisconsin,}\\ 
{\normalsize
\it Madison, WI 53706, USA}
\end{center}
\vspace{0.50cm}
\thispagestyle{empty}
\begin{abstract}
We derive a condition the short distance coefficients governing
$b\to (s,d)\ell^+\ell^-$ transitions
must satisfy in order for the forward-backward
asymmetry to vanish in the exclusive modes $B\to (K^*,\rho)\ell^+\ell^-$. 
This relation, which is satisfied in the standard model, 
involves the coefficient entering in $b\to s\gamma$ transitions as well
as one of the additional Wilson coefficients present in the leptonic modes.
We show that the resulting relation is largely free of hadronic 
uncertainties,
thus constituting a reliable test of the standard model in exclusive rare 
$B$ decays.
\end{abstract}
\newpage

\renewcommand{\thefootnote}{\arabic{footnote}}
\setcounter{footnote}{0}
\setcounter{page}{1}

\section{Introduction} 
\vspace{-0.2cm}
Transitions involving Flavor Changing Neutral Currents (FCNC) have 
attracted a great deal of interest given that they are forbidden 
at tree level in the Standard Model (SM). This suggests  that they have 
a great potential as  tests of the SM as well as to bound its extensions.
This is particularly true of decays governed by the transitions
$b\to q\gamma$, 
$b\to q\ell^+\ell^-$ (q=s,d) and similar other FCNC decays of the $b$
quark. 
It is generally believed that this potential 
is mostly realized in inclusive decays, given that these are
theoretically under control. However, these tend to 
present more experimental difficulties~\cite{bsum}.
On the other hand, 
the predictions for exclusive decay modes are plagued with large
theoretical uncertainties originating in the hadronic matrix
elements. This makes it, a priori, impossible to extract any useful
short distance information from the experimental observations of these
decays. This is certainly the case for $B\to K^*\gamma$. 
Although this also applies to the predictions of the hadronic matrix 
elements in the $b\to q\ell^+\ell^-$ modes,
the combination of symmetries with other experimental
observations can drastically
reduce
the theoretical uncertainties in some decay modes. 
Such is the case for the decay $B\to V\ell^+\ell^-$ ~$(V=K^*,\rho)$, 
for which 
the form-factors can be predicted using a combination of Heavy Quark
Spin Symmetry (HQSS), isospin symmetry ($SU(3)$ for $V=K^*$) 
and the form-factors to be
measured in $B\to\rho\ell\nu$~\cite{wyler,afb}. 
Thus, relatively safe predictions
can be made for the decay rate, 
as well for the forward-backward asymmetry of leptons as a function of
the dilepton mass, $A_{FB}(m_{\ell\ell})$. The latter has been shown
to be very sensitive to extensions of the SM~\cite{afb}. 

In the SM, $A_{FB}(s)$ vanishes for a certain value of $s$. This is
the case in inclusive decays as well as in the exclusive modes
$B\to V\ell^+\ell^-$. 
In this letter we will show that the determination of the dilepton 
mass $s_0$ for which $A_{FB}(s)$ vanishes, constitutes a stringent 
test of the SM even in the exclusive decay modes. 
We will derive a new relation among the short distance Wilson 
coefficients governing the $b\to q\ell^+\ell^-$ transitions,
that results from the vanishing condition for $A_{FB}(s)$, 
and show that this condition is not affected by large theoretical
uncertainties in exclusive channels. 

The separation of short and long distance physics takes place 
in the  operator product expansion
of the effective hamiltonian. This is given by
\begin{equation}
{\cal H}_{\rm eff.}=-\frac{4G_F}{\sqrt{2}}V^*_{tb}V_{tq}
\;\sum_{i}C_i(\mu)O(\mu) ~,
\label{heff}
\end{equation}
where the operator basis $\{O_i\}$ is defined in \cite{heff}, 
$\mu$ is a renormalization scale
and the Wilson coefficient functions $C_i(\mu)$ are determined by the 
short distance structure of the underlying physics. 
To compute the amplitude for the exclusive modes we will need 
the hadronic matrix elements of the operators $O_i$. 
The 
$B\to K^*\ell^+\ell^-$ mode dominates in the SM due to the 
CKM suppression of the $\rho$ mode. 
The main results of this paper are generally valid for any vector
meson $V$. 
The Lorentz structure of the operators defines various form-factors.
The matrix elements necessary to describe this decay are
\begin{eqnarray}
\langle V(k,\epsilon)|\bar{s}_L\gamma_\mu b_L|B({\bf p})\rangle
&=&\frac{1}{2}\left\{ ig\;\epsilon_{\mu\nu\alpha\beta}\epsilon^{*\nu}
(p+k)^\alpha (p-k)^\beta
-f\;\epsilon^*_\mu \right. \nonumber\\
& &\left.- a_+\; (\epsilon^*.p)\;(p+k)_\mu
-a_-\; (\epsilon^*.p)\; (p-k)_\mu\;\right\} ~,
\label{vacur}
\end{eqnarray}
and 
\begin{eqnarray}
\langle V(k,\epsilon)|\bar{s}_L\sigma_{\mu\nu}b_L|B({\bf p}\rangle &=&
\frac{1}{2}\epsilon_{\mu\nu\alpha\beta}\;
\left\{A\;\epsilon^{*\alpha}p^\beta + B\;\epsilon^{*\alpha}k^\beta 
+C\;(\epsilon.p)\;p^\alpha k^\beta\right\}\nonumber\\
& &+\frac{i}{2}\left\{ A\;(\epsilon^{*\mu} p^\nu -\epsilon^{*\nu} p^\mu)
+B\; (\epsilon^{*\mu} k^\nu -\epsilon^{*\nu} k^\mu)\right.\nonumber\\
& &\left.+C\; (\epsilon^*.p)\;(p^\mu k^\nu - p^\nu k^\mu) 
\right\} ~.
\label{sigma}
\end{eqnarray}
In equations (\ref{vacur}) and (\ref{sigma}) the form-factors
$g,~f,~a_{\pm},~A,~B~$ and $C$ are unknown functions of the dilepton
mass squared $s$. In order to compute these, one needs to model the hadron
dynamics involved in the $B\to V$ transition, introducing a large
theoretical uncertainty. 
This obscures the extraction of the interesting short distance
information, encoded in the Wilson coefficients corresponding to the 
operators 
$O_7$, $O_9$ and 
$O_{10}$, which are the relevant ones in $b\to q\ell^+\ell^-$ transitions.   

The forward-backward asymmetry for leptons as a function of the
dilepton mass squared is defined as
\begin{equation}
A_{FB}(s)=\frac{
\int_{0}^{1}\frac{d^2\Gamma}{dxds} dx - 
\int_{-1}^{0}\frac{d^2\Gamma}{dxds} dx 
}
{\frac{d\Gamma}{ds}}  ~,
\label{afbdef}
\end{equation}
where $x=\cos\theta$ and $\theta$ is the angle between the $\ell^+$ 
and the decaying $B$ meson in the $\ell^+\ell^-$ rest frame. 
It is straightforward to show that the numerator of $A_{FB}(s)$ takes
the form
\begin{equation}
A_{FB}(s) \sim  4\;m_B\;k\;C_{10}
\left\{ \bar{C}_9\;g\;f +\frac{m_b}{s}\;\bar{C}_7
\;\left(f\;G - g\;F\right) \right\}~,
\label{numerator}
\end{equation}
where $k$ is the $V$ three-momentum in the $B$ rest frame, 
and we have defined 
\begin{eqnarray}
F&=&A \; p.q + B \; k.q ~, \nonumber\\
G&=&-\frac{(A+B)}{2}~.
\label{fgdef}
\end{eqnarray}
In equation (\ref{numerator}) $\bar{C}_7=C^{\rm
eff.}_7(m_b)$ and  $\bar{C}_9=C^{\rm eff.}_9(m_b)$ are the effective  
Wilson coefficients at the scale $m_b$. These include
all the effects of the renormalization group running as well as, in
the case of $\bar{C}_9$, the long distance effects~\cite{rey} 
associated with off-shell $c\bar c$ intermediate states
\footnote{It is assumed that 
the resonant $J/\psi$ and $\psi'$ contributions are explicitly removed. 
Various treatments of the long distance contributions exist.
The associated uncertainty, however, has very little effect well below
the $J/\psi$, where $s_0$ is likely to be.}.
Thus, $A_{FB}(s)$ vanishes for a value of $s$ determined only by two
of the three Wilson coefficients, $\bar{C}_7$ and  $\bar{C}_9$. 
This condition can be written as the relation
\begin{equation}
\bar{C}_9=-\frac{m_b}{s_0}\;\bar{C}_7
\;\left(\frac{G}{g}-\frac{F}{f}\right)~,
\label{con1}
\end{equation}
where $s_0$ corresponds to the dilepton mass for which $A_{FB}(s)=0$
is satisfied, and the form-factors are evaluated at $s_0$.
The condition (\ref{con1}) for the vanishing of $A_{FB}(s)$ 
constitutes a potentially powerful test of the SM given that it relates 
the Wilson coefficient governing $b\to s\gamma$ decays, $\bar{C}_7$, 
to one of the additional coefficients appearing in the leptonic modes, the 
one that determines the vector coupling to the lepton current.
However, and as it is frequently the case for exclusive decay modes, 
large theoretical uncertainties are present in  (\ref{con1}), a result
of our inability to compute the form-factors $F(s)$, $G(s)$, $f(s)$ and 
$g(s)$ within a controlled approximation.
In what follows we show that, with the use of well established symmetry 
arguments, it is possible to derive from (\ref{con1}) a relation between
$\bar{C}_7$ and $\bar{C}_9$ that is largely free of the hadronic
theoretical uncertainties mentioned above.

In the limit $m_b\gg\Lambda$, with $\Lambda$ the typical scale of the 
strong interactions inside the $B$ meson, the spin of the $b$ quark
decouples from the light degrees of freedom~\cite{iwsym}. 
This results in various
relations among hadronic matrix elements and, therefore, among the 
form-factors parameterizing them. We will refer to these 
as Heavy Quark Spin Symmetry (HQSS) relations. The HQSS relations
corresponding to the matrix elements
of (\ref{vacur}) and (\ref{sigma})  allow us to express the form-factors
$F$ and $G$ as functions of the ``semileptonic'' form-factors
$f$ and $g$~\cite{iw}. They take the form
\begin{eqnarray}
F &=&-f\left(m_B-E_V\right) - 
2m_Bg\left(m_BE_V+k^2\right)~, \label{hqsf}\\
G&=&\frac{f+2m_B\left(m_B-E_V\right)g}{2m_B}~, 
\label{hqsg}
\end{eqnarray}
where $E_V$ and $k$ are  the energy and momentum 
of the $V$ meson in the $B$ rest frame, respectively.
Furthermore, the form-factors $f$ and $g$ entering in the $(V-A)$ 
$B\to V$ matrix element (\ref{vacur}), can be identified with the 
analogous form-factors entering in the semileptonic decay
$B\to\rho\ell\nu$.  
In the case $V=\rho$ this identification only makes use
of isospin symmetry, whereas for $V=K^*$ 
the use of $SU(3)$ symmetry is required. We address 
the issue of $SU(3)$ corrections later in the paper. 
We can then rewrite the condition (\ref{con1}) making use of (\ref{hqsf}) and
(\ref{hqsg}), which  gives
\begin{equation}
\bar{C}_9=-\frac{m_b}{2\;s_0}\;\bar{C}_7
\;\left\{4m_Bk^2\;R_V+\frac{1}{m_B\;R_V}+4(m_B-E_V)\right\}~,
\label{con2}
\end{equation}
where we defined the ratio
\begin{equation}
R_V\equiv \frac{g(s_0)}{f(s_0)}~,
\label{rvratio}
\end{equation}
and all quantities depending on the dilepton mass must be evaluated at
$s=s_0$. This is the main result of the paper. The relation (\ref{con2})
between $\bar{C}_7$ and $\bar{C}_9$ now only depends on the ratio of the 
vector to axial-vector form-factors $R_V$, which in turn can be 
experimentally extracted from the decay $B\to\rho\ell\nu$. 

Corrections to (\ref{con2}) are expected to be small. 
The HQSS relations (\ref{hqsf}) and 
(\ref{hqsg}) receive corrections suppressed by inverse powers
of the $b$ quark mass. These come from the fact that the HQSS neglects 
the lower components of the $b$-quark spinor. Thus, the suppressed terms are
proportional to $p_b/m_b$, where $p_b$ is the $b$ quark momentum in the $B$
meson rest frame and is of the order of the typical momentum
exchanged with the light degrees of freedom, $\Lambda$. 
Thus, we expect the typical size of these corrections to be 
of the order of $10\%$ or less. 

Up to this point we have not specified the vector meson in the
final state. In the SM, the branching ratio for the 
$K^*$ mode is expected to be 
about a factor of $20$ larger than the one for the $\rho$ mode, due to 
the ratio of CKM matrix elements $V_{ts}/V_{td}$. 
For the $B\to K^*\ell^+\ell^-$ decay is important to address the    
corrections to the $SU(3)$ identification
of the form-factors $f$ and $g$ with those entering in 
the semileptonic decay  $B\to\rho\ell\nu$. Estimates of these corrections
in specific calculations indicate they are small as long as
the recoil energy of the $K^*$ is large enough. To show this explicitly 
we can make use of some general properties of the 
constituent quark model picture that are likely to capture 
the correct $SU(3)$ breaking effects. 
This is the situation if one uses the formalism proposed by 
Stech~\cite{stech} and further developed by Soares~\cite{joao} to 
include light quark mass effects, 
which is fully relativistic and incorporates
correctly the quark spin degrees of freedom. The spin structure
plays a fundamental role in the ratios of form-factors. 
We are interested in estimating the double ratio
\begin{equation}
\delta\equiv \frac{R_V^{K^*}}{R_V^{\rho}}~,
\label{delta}
\end{equation}
where $R_V^{K^*}$ refers to the quantity in (\ref{con2}), whereas
$R_V^\rho$ corresponds to the quantity extracted from $B\to\rho\ell\nu$.
The deviations from $\delta=1$ are a measure of the amount of $SU(3)$ 
breaking. 
Within the formalism of Reference~\cite{stech,joao} we obtain the approximate
expression
\begin{equation}
\delta\simeq \frac{1+m_d/(E_{K^*}-E_{\rm sp})}
{1+m_s/(E_{K^*}-E_{\rm sp})}~,
\label{estdel}
\end{equation}
where $m_d$ and $m_s$ are the down and strange quark constituent masses, 
and $E_{\rm sp}$ is the energy of the spectator quark inside the 
vector meson. This is typically of the order of $\Lambda$, i.e. a few
hundred MeV. Thus, for large enough values of $E_{K^*}$, 
the ratio $R_V$ is not 
very sensitive to $SU(3)$ breaking effects. For instance, for the typical 
values $m_d=E_{\rm sp}=300~$MeV, $m_s=450~$MeV, the $SU(3)$ breaking effect
is below $15\%$ for $E_{K^*}>1~$GeV. As we will see below,  
the typical recoil energies where the asymmetry vanishes are even larger. 

The measurement of the ratio $R_V$ from $B\to\rho\ell\nu$
decays will hopefully be available in the $B$ factory era. 
Thus, the measurement of $s_0$ in any of the 
$B\to V\ell^+\ell^-$ modes can be turn into a test of the SM via the relation
(\ref{con2}).   
However, it is interesting to estimate the value of $s_0$ in the SM, 
in order see that it typically 
corresponds to a region with large recoil energy $E_V$. 
In order to illustrate this point we can use again Stech's formalism. 
Then the ratio of vector to axial-vector form-factors is simply  
\begin{equation}
R_V\simeq -\frac{1}{2\;m_B\;k}~. 
\label{steap}
\end{equation}
Now the 
condition (\ref{con2}) for the vanishing of the asymmetry 
simplifies to
\begin{equation}
\bar{C}_9\simeq -2\;\frac{m_b}{s_0}\;\bar{C}_7\;(m_B-E_V-k)~,
\label{con3}
\end{equation}
which, solving for $s_0$, translates into 
\begin{equation}
s_0\simeq\frac{m_B^2+m_V^2\left(2\bar{C}_7/\bar{C}_9-1\right)}
{1-\bar{C}_9/2\bar{C}_7}~.
\label{snot}
\end{equation}
We can use this expression to obtain an estimate of $s_0$ in the 
SM. For instance, using the next-to-leading order value for $\bar{C}_7$
\cite{misiak} and the corresponding value of $\bar{C}_9$ as described in
\cite{c9} one obtains, for $V=K^*$, 
$s_0\simeq 3.9~{\rm GeV}^2$. This value is in remarkable
agreement with what it is obtained with typical model
calculations of $R_V$. 
This is not entirely surprising since, although Stech's formalism makes use
of the constituent quark picture, the ratio $R_V$ is independent of 
wave-functions and overlap integrals, which typically are the main source of 
disagreement among different calculations of individual form-factors. 
In Figure~1 we illustrate this point by plotting the non-resonant
forward-backward asymmetry
$A_{FB}(s)$ defined in (\ref{afbdef}) as a function of the dilepton mass
$s$, for the model calculations of References~\cite{bsw,ball,meli}. 
The location of the zero of
the lepton asymmetry is fully determined by $R_V$. This ratio tends  to be 
very similar across models, even when the values of the individual 
form-factors may differ. Also shown, is the result of one of the 
models (BSW*) obtained by significantly changing $R_V$ by
doubling the value of the vector form-factor $g$. 
The resulting shift in the position of the asymmetry zero 
gives a conservative estimate of the theoretical uncertainty one incurs
in by using models. On the other hand,  
such shift in $R_V$ would significantly affect the $B\to\rho\ell\nu$
branching ratio, enhancing it by a factor of $(2-3)$ depending on the 
$s$ dependence. Such dramatic effects, already bound by the present
CLEO measurement of this mode~\cite{cleo}, 
will be extremely constrained by more precise measurements in the 
$B$-factory era.
Thus, we conclude
that the value of $s_0$ is much less sensitive to changes in $R_V$
than $B\to\rho\ell\nu$. In this way, we see that  high precision 
in the extraction of $R_V$ {\em is not a necessary condition}
in order to have a precise prediction of the position of the zero
in the asymmetry $A_{FB}(s)$.
 
Extensions of the SM modify the matching conditions of 
the short distance coefficients~\cite{afb,bsm},
therefore
potentially upsetting the relation (\ref{con2}). 
A change in $\bar{C}_7$ and/or 
$\bar{C}_9$ would appear as a shift in $s_0$. On the other hand, a new 
contribution mainly affecting $C_{10}$ would have no effect on the zero of
$A_{FB}(s)$, whereas it would affect other quantities such as momentum
distributions, branching ratios, etc. Such is the discriminating power of 
measuring the location of the $A_{FB}(s)$ zero. 
For instance, since the sign of $\bar{C}_7$ is not measured
in $b\to s\gamma$, it is in principle possible that it is the opposite
to the SM prediction. In this extreme case, the forward-backward asymmetry
does not have a zero in the physical region. Less drastic modifications
occur in several scenarios involving new states which contribute to the 
one-loop $b\to q\ell^+\ell^-$ transition amplitude. 

The current experimental limits~\cite{exp} on $b\to q\ell^+\ell^-$ processes, 
although still above the SM expectations, indicate that sensitivity to 
these transitions will be achieved soon and that, in some cases, large data
samples could be accumulated in the near future. 
We have shown that it is possible to reliably test the SM in exclusive
FCNC $B$ decays. In particular, we have 
seen that the measurement of the zero of the forward-backward asymmetry
for leptons, $A_{FB}(s)$, in $B\to V\ell^+\ell^-$ decays 
provides a test of the short distance structure
of the SM and its extensions, within a controlled approximation. 
The relation (\ref{con2}) involving the Wilson coefficients $\bar{C}_7$
and $\bar{C}_9$ is derived by making use of the heavy quark spin symmetry, 
and is expected to receive only small corrections. These are the same 
corrections leading to $(m_{B^*}-m_B)/m_B~\simeq 0.009$. 
The experimental measurement of the ratio of form-factors $R_V$ 
from $B\to\rho\ell\nu$ decays, even if not a very precise one,
together with the condition (\ref{con2}), provides
a stringent test of the SM in the 
CKM-suppressed mode $B\to\rho\ell^+\ell^-$.
The CKM-favored mode
$B\to K^*\ell^+\ell^-$ requires the use of $SU(3)$ symmetry relations
among the form-factors. We estimated the $SU(3)$ breaking corrections in 
(\ref{estdel}) to be small for a fast recoiling $K^*$. On the other hand, 
we have also estimated the approximate value of the dilepton mass $s_0$
for which $A_{FB}$ vanishes and found it to be typically
at a lepton mass corresponding
to $E_{K^*}\simeq 2.3~$GeV which, according to (\ref{estdel}), would
imply a very small $SU(3)$ correction of the order of $6\%$.
Thus this exclusive mode, which is experimentally favored over  
other exclusive channels as well as over the inclusive decay, 
provides a test of the short distance structure of flavor changing
neutral currents. 

\vskip1.50cm
\noindent
{\bf Acknowledgments}

\noindent
The author thanks P. Ball for providing a parametrization of the model
of reference~\cite{ball}, L. Reina for a discussion on the next-to-leading
order calculation of $\bar{C}_7$, and J. D. Richman for discussions 
concerning the experimental prospects of $B\to\rho\ell\nu$ measurements.
This work was supported  by the U.S.~Department of Energy under  
Grant No.~DE-FG02-95ER40896 and the University of 
Wisconsin Research Committee with funds granted by the Wisconsin 
Alumni Research Foundation.


\vskip 3cm
\begin{center}
\large\bf Figure Caption
\end{center}

\noindent
The non-resonant forward-backward asymmetry of leptons $A_{FB}$ 
defined in (\ref{afbdef}), for 
$B\to K^* e^+ e^-$ as a function of the dilepton mass $s$. 
The asymmetry is computed by making use of the relations~(\ref{hqsf})
and (\ref{hqsg}) and the semileptonic form-factors from: the BSW* model
of reference~\cite{bsw} (solid line), the light-cone QCD sum rule calculation
of reference~\cite{ball} (dashed line) and the relativistic quark model of 
reference~\cite{meli} (dotted line).  The lighter solid line 
corresponds
to the BSW* model with the vector form-factor $g$ multiplied by a factor 
of two, and illustrates the uncertainty in the position of the $A_{FB}$ zero. 

\vskip 0.5cm

\newpage
\begin{figure}[p]
\center
\hspace*{-1.0cm}
\psfig{figure=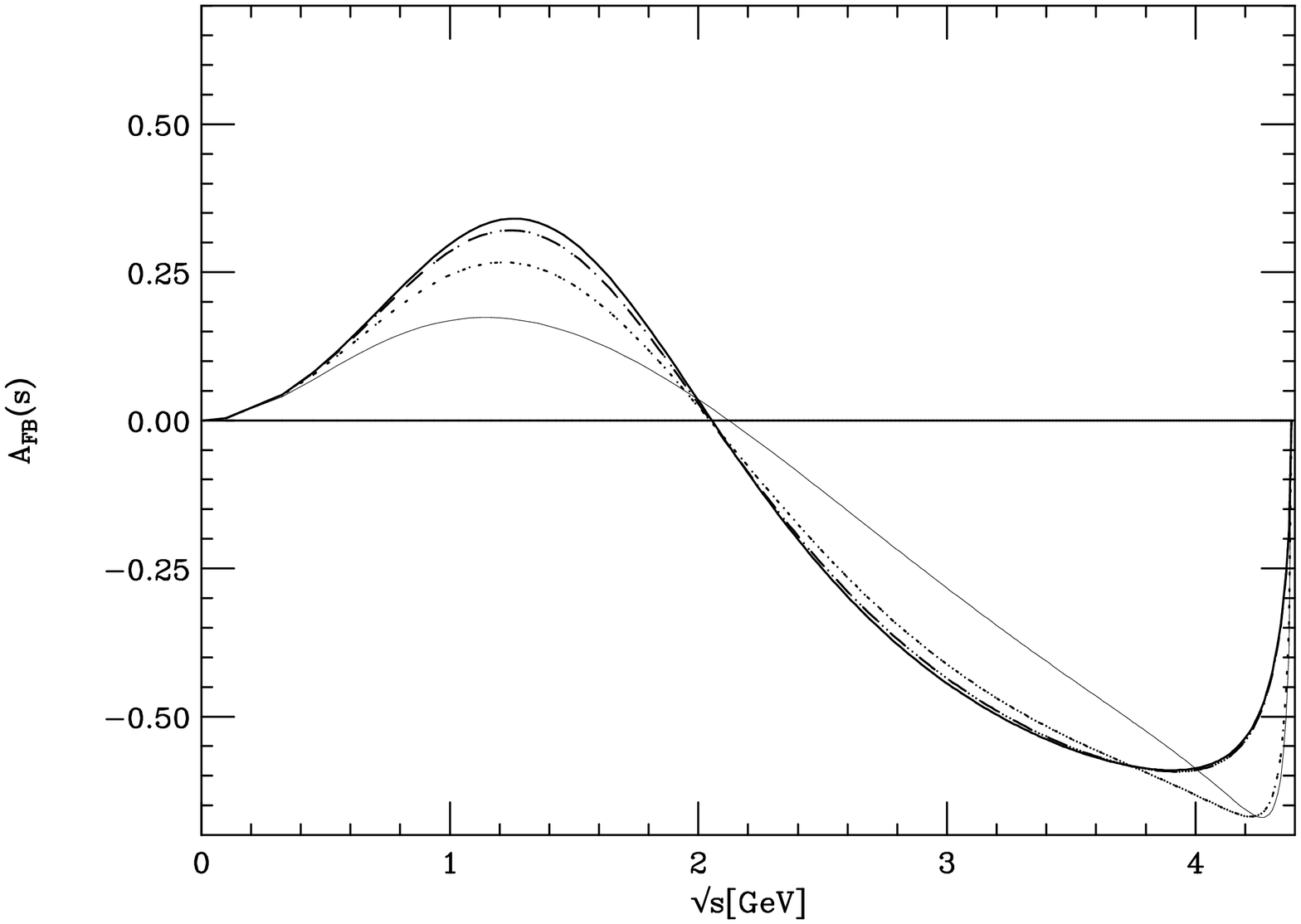,height=5in,angle=90}
\vspace*{2cm}
\end{figure}

\end{document}